\documentclass[journal,draftclsnofoot,onecolumn,12pt]{IEEEtran}
\usepackage{cite}

\usepackage{verbatim}
\ifCLASSINFOpdf

\else

\fi
\usepackage{array}
\usepackage{algorithm}
\usepackage{algorithmic}
\usepackage{amsmath}
\input{amssym}

\usepackage{float}
\usepackage{array}
\usepackage{psfrag}
\usepackage{balance,multirow}
\usepackage{hyperref}
\usepackage{graphicx}
\usepackage{caption}
\usepackage{booktabs}
\usepackage{subcaption}
\usepackage{booktabs}
\usepackage{breakurl}
\usepackage{tikz}

\usetikzlibrary{arrows,positioning,decorations.markings}

\newcolumntype{C}[1]{>{\centering\let\newline\\\arraybackslash\hspace{0pt}}m{#1}}

\newtheorem{prop}{Proposition}

\begin{document}

\title{Adaptive Delivery in Caching Networks }

\author{Seyed Ali Saberali, Hamidreza Ebrahimzadeh Saffar, Lutz Lampe, and Ian Blake\\
}

\maketitle
\begin{abstract}
The problem of content delivery in caching networks is investigated for scenarios where multiple users request identical files. Redundant user demands are likely when the file popularity distribution is highly non-uniform or the user demands are positively correlated.
An adaptive method is proposed for the delivery of redundant demands in caching networks. Based on the redundancy pattern in the current demand vector, the proposed method decides between the transmission of uncoded messages or the coded messages of \cite{Maddah_decentralized:2014} for delivery. Moreover, a lower bound on the delivery rate of redundant requests is derived based on a cutset bound argument. 
The performance of the adaptive method is investigated through numerical examples of the delivery rate of several specific demand vectors as well as the average delivery rate of a caching network with correlated requests. The adaptive method is shown to considerably reduce the gap between the non-adaptive delivery rate and the lower bound.  In some specific cases, using the adaptive method,  this gap shrinks by almost 50\% for the average rate. 
\end{abstract}
\begin{IEEEkeywords}
Adaptive delivery algorithm, average delivery rate, coded caching, correlated requests, placement optimization, redundant demands.
\end{IEEEkeywords}
\section{Introduction}\label{sec_int}
\IEEEPARstart{L}{ocal} content caching is a promising technique to meet the unprecedented traffic demands in the next generation communication networks \cite{Maddah_decentralized:2014,Maddah_limits:2014,5gMagazine:2014,Golrezaei:2013,Caire:2014,Digavi:2015}. Caching networks take advantage of the users' contextual information to  predict the future user demands. This enables the network to store the popular content at storage nodes, also known as caches, close to the end users and satisfy the user requests locally \cite{5gMagazine:2014}. 

Caching networks operate in two phases, which are commonly referred to as placement and delivery phases. In the placement phase, the caches fill their memories with parts of the popular files up to their storage capacity. This phase takes place when the network traffic is low. In contrast, the delivery phase is performed when the network is congested. In this phase, upon the users' requests, each cache provides its users with the parts of the files that it has available. The remaining parts of the files are conventionally delivered to the users  through separate unicast transmissions performed by a central server on a channel that is shared by the users. 
In a more recent caching approach \cite{Maddah_limits:2014}, known as coded caching, the central server uses simultaneous coded-multicasting to deliver the requested content to the users to further reduce the network congestion. 

An information-theoretic formulation of coded caching was developed in \cite{Maddah_limits:2014}. 
The authors defined the delivery rate as the total traffic on the shared communication link due to the server's messages, such that the users' requests are satisfied. Moreover, they proposed a centralized coded-caching scheme to reduce the delivery rate.  
In a later work \cite{Maddah_decentralized:2014}, a decentralized caching scheme was proposed that did not require any coordination between the caches to operate. 
The decentralized nature of this method made it the building block of several caching schemes that were designed later for  more complicated scenarios \cite{Maddah_nonuniform:2014,Zhang:2015,Digavi:2015,maddah_Online:2014,Digavi2:2015,Maddah_hier:2014}.

Both \cite{Maddah_decentralized:2014,Maddah_limits:2014} used the peak delivery rate as the figure of merit of the caching network. The peak rate results from the worst-case demand vector, where all the users request distinct files. However, the average delivery rate is also a significant performance metric of a caching network. 
Average delivery rate depends on the statistics of the user requests. Thus, the statistical patterns in the user demands can significantly affect the design of the caching scheme. 

One  statistical property of user demands is the popularity distribution of the files.
The caching schemes of \cite{Maddah_decentralized:2014,Maddah_limits:2014} can be used if the popularity distribution is uniform.
On the other hand,
\cite{Maddah_nonuniform:2014,Digavi:2015,Zhang:2015,Caire:2014} have proposed different caching schemes to account for non-uniform popularities. In particular, the caching schemes of \cite{Maddah_nonuniform:2014,Digavi:2015} are designed based on grouping of the files into several popularity groups, with the files in each group having relatively close popularity levels. They provide more storage resources to the files in the more popular groups. Then, they use the decentralized caching scheme of \cite{Maddah_decentralized:2014} within each group separately. Also, \cite{Zhang:2015} groups the library of files into two groups of popular and unpopular files. The requests for popular files are delivered through the delivery algorithm of \cite{Maddah_decentralized:2014}, while the requests of unpopular files are delivered through  uncoded messages. The same problem is investigated in \cite{Caire:2014}, assuming a Zipf popularity distribution and  independent and identically distributed user requests. The placement of \cite{Caire:2014} is based on the partitioning of each file into equal length packets and randomly distributing the packets (not bits) over the caches. Unlike the other schemes, the delivery of \cite{Caire:2014} is not based on the delivery algorithm of \cite{Maddah_decentralized:2014}, but on chromatic number index coding. In contrast to \cite{Maddah_nonuniform:2014,Digavi:2015,Zhang:2015}, this scheme does not restrict the coding opportunities to the requests within each popularity group. However, its implementation is more complicated as it requires vertex coloring of a conflict graph.

The statistics of  the users' requests can further affect the design of caching networks through increasing the chance of multiple identical requests. In such a scenario, one might be able to modify the delivery algorithm to benefit from the redundancies in the user demands, to further reduce the average delivery rate.
Redundant demands are likely to be made when the files have significantly different popularity levels or when there are positive correlations among the requests of different users.
For the case of non-uniform file popularities, the schemes in \cite{Maddah_nonuniform:2014,Digavi:2015,Zhang:2015} do not take the effect of identical requests into account during the delivery phase. This is because the delivery in all these schemes is based on the delivery of \cite{Maddah_decentralized:2014}, which is designed for the demand vectors with distinct requests.  
In addition to non-uniform popularity levels, correlated user requests are likely in many practical scenarios. 
A considerable amount of multimedia requests are made through the social networks like Facebook, Twitter and Instagram and movie providing websites like Netflix. In such scenarios, the users with overlapping circles of friends, the ones who follow the same people or pages, and those  who live in the same geographical area or have common personal, social and professional interests are likely to get suggestions for the same content in their media feeds, and therefore, request the same files. 

In this paper, we investigate the delivery of redundant demands in caching networks.
We study a model where placement is fixed, yet the requests are changing by the time and the delivery adapts to the requests. We propose an adaptive delivery scheme based on \textit{message selection} to minimize the delivery traffic. Specifically, upon receiving a demand vector from the users, the server exploits the redundancy pattern in the user demands to decide whether to use uncoded messages or the coded messages of \cite{Maddah_decentralized:2014}  to deliver each part of the files requested. 

We  assume that the placement phase is accomplished through the placement schemes of \cite{Maddah_limits:2014,Maddah_decentralized:2014}. 
This ensures that the peak delivery rate does not exceed the delivery rates of \cite{Maddah_limits:2014,Maddah_decentralized:2014}, so the link capacity constraints are satisfied. Further, if the file popularities are relatively uniform or little prior knowledge about the popularity distribution is available during the placement time, it is natural to accomplish the placement as in \cite{Maddah_limits:2014,Maddah_decentralized:2014}.
In the delivery phase, however, the users reveal their demands to the server. The server can use this knowledge as a side information and  adapt its choice of coded and uncoded messages accordingly to benefit from the possible redundancies in the  requests. To the best of the authors' knowledge, this paper is the first work in the literature to consider this scenario and to specifically design a scheme for the delivery of redundant requests. 

Although we use the placement schemes in \cite{Maddah_limits:2014,Maddah_decentralized:2014}, our proposed delivery method is based on an optimization formulation of the content placement problem. Namely, we use a modified version of this problem to optimize the choice of coded and uncoded messages in our proposed delivery scheme. A side result of the placement optimization problem is the generalization of the centralized placement of \cite{Maddah_limits:2014} to arbitrary cache sizes. In particular, we derive the parameters of the  centralized caching analytically for cases where the total cache capacity is not an integer multiple of the total size of the files in the library.

We show the superiority of our adaptive method through  numerical examples for several specific demand vectors. We derive a lower bound on the delivery rate of the redundant requests based on a cutset bound argument, and compare the rate of the proposed  delivery method with the lower bound. Moreover, we study the dynamics of a caching system with correlated user demands. We apply Gibbs sampling \cite{Murphy:2012,Fischer:2012}, to generate sample demand vectors based on a stochastic modeling of the dependencies among the user requests. It is shown that the proposed method is superior to the conventional non-adaptive method in terms of the average delivery rate. In some specific cases, the adaptive method decreases the gap between the average rate of the non-adaptive scheme and the lower bound by almost 50\%.

The remainder of this paper is organized as follows. In Sec.~\ref{sec_rev}, we present the network model and review the caching schemes of \cite{Maddah_decentralized:2014,Maddah_limits:2014}. We formulate the rate minimization problem in Sec.~\ref{sec_opt}.  In Sec.~\ref{sec_ada}, we propose the adaptive delivery scheme and derive a lower bound on the delivery rate. Sec.~\ref{sec_sim} presents numerical examples and simulation results. Finally, we conclude the paper in Sec.~\ref{sec_con}.

\section{Problem Model and Review}\label{sec_rev}
\tikzset{
  treenode/.style = {align=center, inner sep=0pt,text centered },
  arn_r/.style = {treenode, draw=none, rectangle,rounded corners=1mm,
    fill=white, text width=4.7em,text height=1.7em,text depth=4em},
   arn_r_ser/.style = {treenode, draw=none, rectangle,rounded corners=1mm,
    fill=white, text width=4.7em,text height=1.6em,text depth=5em},
  arn_c/.style = {treenode, circle, black,draw=none,
    fill=white, text width=2em},
  arn_x/.style = {treenode, rectangle, draw=none,
    minimum width=0.0em, minimum height=0.0em}
}
In this section, we explain the problem model and briefly review the caching schemes of \cite{Maddah_decentralized:2014,Maddah_limits:2014}.

\begin{figure}
\centering
\begin{tikzpicture}[>=stealth',level/.style={sibling distance = 4cm/#1,
  level distance = 1.5cm}] 
\node [arn_r_ser] {Server\\\medskip\includegraphics[scale=.15]{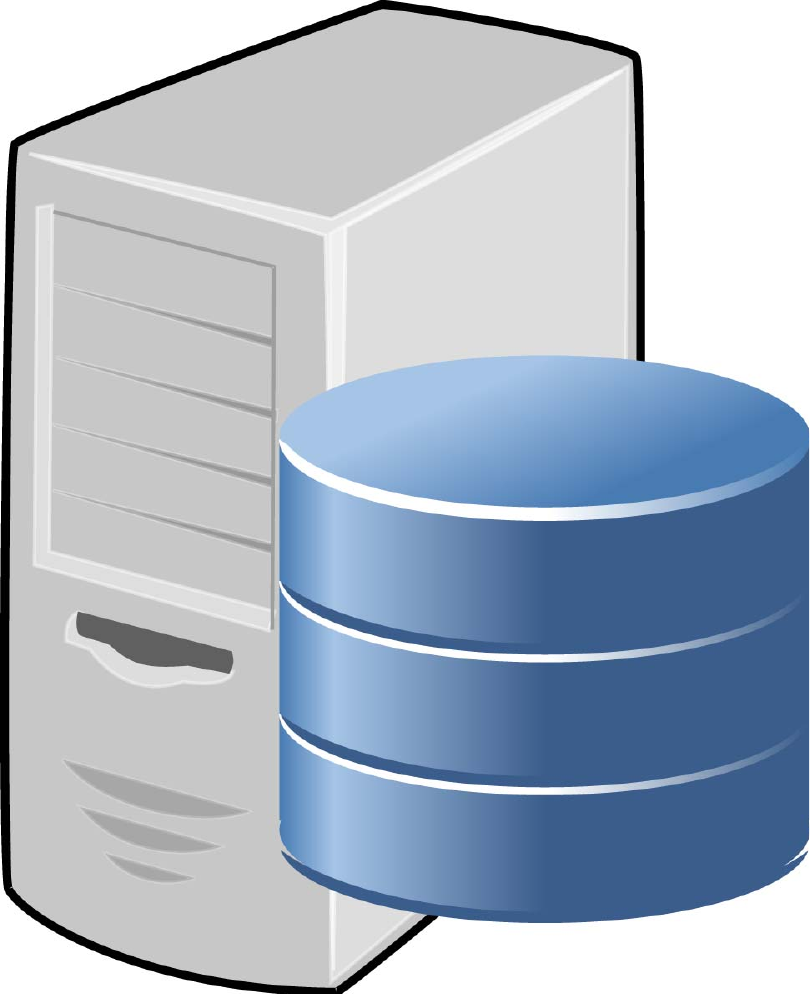}}
    child{ node [arn_x] {}[level/.style={sibling distance = 2cm,level distance = .7cm}] 
            child{ node [arn_x]{} [level/.style={level distance = 1.5cm}] 
            		child{node [arn_r] {\small{Cache $1$}\\\medskip
\includegraphics[scale=0.65]{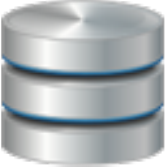}
            		} [level/.style={level distance = 2cm},->] 
            			child{node[arn_c] {\includegraphics[scale=0.14]{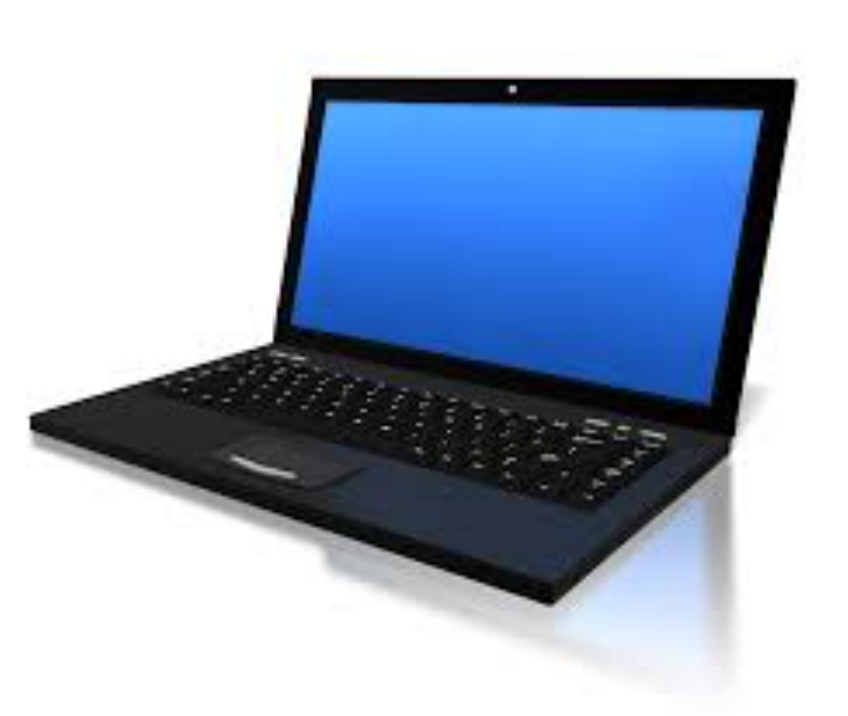}}
            			}
            		}
            }
            child{ node [arn_x]{} [level/.style={level distance = 1.5cm}]
            		child{node [arn_r] (a) {\small{Cache $2$}\\\medskip
\includegraphics[scale=0.65]{database.pdf}
            		}[level/.style={level distance = 2cm},->] 
            			child{node[arn_c] (aa) {\includegraphics[scale=0.14]{laptop.pdf}} 
            			}
            		}
            }
            child{ node [arn_x]{} [level/.style={level distance = 1.5cm}]
            		child{node [arn_r] (b) {\small{Cache $K\!-\!1$}\\\medskip
\includegraphics[scale=0.65]{database.pdf}
            		}[level/.style={level distance = 2cm},->]  
            			child{node[arn_c] (bb) {\includegraphics[scale=0.14]{laptop.pdf}} 
            			}
            		}
            }
            child{ node [arn_x]{} [level/.style={level distance = 1.5cm}]
            		child{node [arn_r]{\small{Cache $K$}\\\medskip
\includegraphics[scale=0.65]{database.pdf}
            		}[level/.style={level distance = 2cm},->]  
            			child{node[arn_c] {\includegraphics[scale=0.14]{laptop.pdf}} 
            			}
            		}
            }                                          
        }
;        
\path (a) -- (b) node [midway] {$\cdots$};      
\end{tikzpicture}
\vspace*{.1cm}
\caption{\small{A network with $K$ caches and a central server. }
}\label{config_fig}
\end{figure}

Assume a network with a central server and $K$ caches, where the server is able to communicate with the caches through a broadcast link (see Fig.~\ref{config_fig}). We denote the set of all  caches in the network by $\mathcal{K}$. A library of $N\geq K$ popular files is given, where each file is $F$ bits long. We assume that all files are available at the central server and that each cache has a memory capacity of $M\times F$ bits.  $q\triangleq  M/N$ represents the ratio of the cache size to the library size. 

\paragraph*{Placement Phase} In the placement phase, the caches fill their memories with parts of the popular files based on a placement algorithm. We assume that placement takes place only once and remains unchanged during the delivery phase.  

The resulting distribution of  bits in the caches can be described as follows. 
For a given file $n$ and a given subset of caches $\mathcal{S}\subset \mathcal{K}$, denote by $V_{\mathcal{S}}^n$ the subset of bits  of file $n$ that are exclusively stored at the caches in $\mathcal{S}$. Note that the resulting subsets of  bits \textit{partition} the set of all the bits of every file into $2^K$ partitions.  Define $s\triangleq |\mathcal{S}|$ and 
\begin{align}
x_s\triangleq |V_{\mathcal{S}}^n|/F
\end{align} 
as the portion of the bits of file $n$ that are exclusively stored at each subset $\mathcal{S}$ of caches with cardinality $s$. Here, we have assumed that $|V_{\mathcal{S}}^n|$ only depends on $s$. In particular, it neither depends on $n$ nor on the particular choice of caches in $\mathcal{S}$ as long as the cardinality of $\mathcal{S}$ is $s$. This holds because of symmetry, as we assume a uniform  distribution over file popularities. 

The placement phase can be performed through either the centralized scheme of \cite{Maddah_limits:2014} or the decentralized scheme of  \cite{Maddah_decentralized:2014}.   
The centralized caching scheme of \cite{Maddah_limits:2014} can be used only when $t\triangleq \frac{KM}{N}$ is an integer.
For the centralized placement, split each file into $\binom{K}{t}$ non-overlapping subfiles of the same length $F/\binom{K}{t}$. Assign each one of these subfiles to a subset of caches $\mathcal{S}:s=t$, in a one-to-one manner. Store the bits belonging to each subfile in all the caches in the corresponding $\mathcal{S}$. This results in 
\begin{align}\label{eq_cen_plc}
    x_s^{\text{cen}} = \begin{cases}
               1/\binom{K}{s}               & s=t\\
               0               & s\neq t
           \end{cases}.
\end{align}
For the decentralized placement, each cache stores $\frac{M}{N}F$ bits of each file uniformly at random. It can be shown that for large $F$  \cite{Maddah_decentralized:2014}
\begin{align}\label{eq_dec_plc}
x_s^{\text{decen}}\approx q^{s-1}(1-q)^{K-s+1},\quad s=0,...,K
\end{align}
with high probability.

\paragraph*{Delivery Phase}  
In the delivery phase, the network serves one user of \textit{every} cache at a time. Denote the requests of the users of caches $1,...,K$  with $d_1,...,d_K$, respectively. We refer to the vector $[d_1,...,d_K]$ as the demand vector. Note that the demand vector evolves with time during the delivery phase. We represent the number of distinct files in the demand vector by $L$, where $1\leq L \leq K$. We call the demand vector \textit{redundant} if $L<K$. In addition, denote by $k_i$, the number of requests for the $i$-th most requested file in the current demand vector. Thus $k_i\geq k_j$ for $ i>j$ and $i,j\in\{1,...,L\}$. We call $(k_1,...,k_L)$ the redundancy pattern of the demand vector. For a demand vector $[d_1,...,d_K]$, we define the delivery rate $R(M,[d_1,...,d_K])$ as the traffic on the shared broadcast link due to the server's messages,  such that all the caches successfully recover the files they requested. We express the rate in terms of the equivalent total number of files that must be transferred on the shared link. So, a rate of $R$ files is equivalent to $R\times F$ bits.

To construct file $d_k$, cache $k$ needs to receive $V^{d_k}_{\mathcal{S}}$ for all $\mathcal{S}\subset\mathcal{K}\backslash\{k\}$. The server, delivers these bits to the caches through the coded delivery messages given by Algorithm~\ref{alg_del0d} proposed in \cite{Maddah_decentralized:2014}. Notice that the delivery method for the centralized caching in \cite{Maddah_limits:2014} is a special case of Algorithm~\ref{alg_del0d}.
\begin{algorithm}                      
\caption{\small{Delivery algorithm of \cite{Maddah_decentralized:2014}}}          
\label{alg_del0d}                           
\begin{algorithmic}                    
\REQUIRE $\{V^{n}_{\mathcal{S}}\}_{ n=1,...,N,\,\mathcal{S}\subset \mathcal{K}}$\quad \textit{\# From the placement phase}
    \STATE  \textbf{Procedure} Delivery($d_1,...,d_K$)
    \FOR{$s=K,K-1,...,1$}
    \FOR{$\mathcal{S}\subset \mathcal{K}:|\mathcal{S}|=s$}
        \STATE server sends $\oplus_{k\in\mathcal{S}}V^{d_k}_{\mathcal{S}\backslash\{k\}}$
    \ENDFOR
    \ENDFOR
\end{algorithmic}
\end{algorithm}
\\
Note that if file $n$ is requested by multiple users, including user $k$, Algorithm~\ref{alg_del0d} embeds $V^n_{\mathcal{S}\backslash\{k\}}$ into several messages. If $s>1$, user $k$ has the side information to  decode only one of those messages. As a result,  the server needs to send all the messages with $s>1$, even though the demand vector is redundant. This is not the case for the messages with $s=1$, i.e., $\mathcal{S}=\{k\}$. In these cases, $\oplus_{k\in\mathcal{S}}V^{d_k}_{\mathcal{S}\backslash\{k\}}=V^{d_k}_{\emptyset}$. Such uncoded messages deliver the bits that are not stored at any cache in the system. All the users that request file $n$  can decode $V^{n}_{\emptyset}$, so it needs to be sent only once.  As a result, the traffic due to the uncoded messages is $Lx_0$ instead of $Kx_0$. Thus, the total delivery rate will be
\begin{align}\label{eq_rate_L}
L x_0+\sum_{s=1}^{K-1}\binom{K}{s+1}x_s. 
\end{align} 
Note that when $L=K$, substitution of (\ref{eq_cen_plc}) and (\ref{eq_dec_plc}) in (\ref{eq_rate_L}) gives
\begin{align}\label{eq_rate_cen}
K\left(1-\frac{M}{N}\right)\frac{1}{1+K\frac{M}{N}}
\end{align}
as the \textit{peak} rate of the centralized caching scheme \cite{Maddah_limits:2014}, and 
\begin{align}\label{eq_rate_decen}
K\left(1-\frac{M}{N}\right)\frac{(1-(1-M/N))^K}{KM/N}
\end{align}
as the \textit{peak} rate of the decentralized caching scheme \cite{Maddah_decentralized:2014}. From (\ref{eq_rate_L}), one notes that for the redundant demand vectors, the actual rate of Algorithm~\ref{alg_del0d} is smaller than  (\ref{eq_rate_cen}) and (\ref{eq_rate_decen}) for the centralized and decentralized caching schemes, respectively. This observation is the basis of our analysis in Sec.~\ref{sec_ada}.

\section{Optimality of the Centralized Placement with delivery based on  Algorithm~\ref{alg_del0d}}\label{sec_opt}
As was formerly discussed, we use either of the methods in \cite{Maddah_decentralized:2014} or \cite{Maddah_limits:2014} for the placement phase of our caching scheme.
In this section, we show that the centralized placement scheme of \cite{Maddah_limits:2014} is the optimal placement to minimize the peak rate of delivery Algorithm~\ref{alg_del0d}. We also generalize the application of the centralized placement to the cases that $t=KM/N$ is not an integer. 

The optimal placement is characterized by the optimal parameters $x_s$ that lead to the smallest peak delivery rate of Algorithm~\ref{alg_del0d}. 
Based on (\ref{eq_rate_L}), the peak rate minimization problem can be formulated as
\begin{align}\label{eq_opt}
\begin{aligned}
& \underset{x_s\textbf{•}}{\text{minimize}}
& & \sum_{s=0}^{K-1}\binom{K}{s+1}x_s\\
& \text{subject to}
& & \sum_{s=0}^{K}\binom{K}{s}x_s=1\\
&&& \sum_{s=1}^{K}\binom{K-1}{s-1}x_s\leq \frac{M}{N}\\
&&& x_s \geq 0,\; s=0,1,...,K.\\
\end{aligned}
\end{align}
The first constraint in (\ref{eq_opt}) ensures that the resulting  subsets partition the bits of each file. It also guarantees that $x_s\leq 1$. The second constraint represents the storage capacity constraint. The objective function of (\ref{eq_opt}) is the the worst-case delivery rate of Algorithm~\ref{alg_del0d}. 
We now present the analytical solution of (\ref{eq_opt}) in Proposition~\ref{lem_plc}.
\begin{prop}[Optimal Placement for the Worst-Case Demand]\label{lem_plc}
Let $t=K\frac{M}{N}$. The solution of (\ref{eq_opt}) is  
\begin{subequations}\label{eq_optplc}
\begin{align}\label{eq_optplc_1x}
x^*_s= \begin{cases}
               1/{\binom{K}{t}},               & s = t\\
               0,               & \text{otherwise}\\
           \end{cases}
\end{align}
if $t$ is an integer, and is
\begin{align}\label{eq_optplc_2x}
x^*_s= \begin{cases}
               ({\lceil t \rceil}-t)/{\binom{K}{{\lfloor t \rfloor}}},            & s = \lfloor t \rfloor\\
              (t-{\lfloor t \rfloor})/{\binom{K}{{\lceil t \rceil}}},           & s = \lceil t \rceil\\
               0,               & \text{otherwise}
           \end{cases}
\end{align}
\end{subequations}
if $t$ is not an integer. ${\lfloor t \rfloor}$ and ${\lceil t \rceil}$ denote the largest integer smaller than $t$ and the smallest integer larger than $t$, respectively.
\end{prop}
\textit{Proof.} See Appendix~\ref{app_plc}.\\
Proposition~\ref{lem_plc} shows that the centralized placement of \cite{Maddah_limits:2014}  is optimal for Algorithm~\ref{alg_del0d} when $t$ is an integer. Further, it generalizes the centralized placement scheme to the caching systems with non-integer $t$.
Table~\ref{tab_sol} shows the optimal placement parameters for a system of $K=5$ caches and a library of $N=1000$ files for various storage capacities. Note that two $x^*_s$ values are non-zero when $t$ is non-integer.
\begin{table}
\centering
\begin{tabular}{c|cccccc|C{2cm}} 
$M/N$  & $x_0^*$ & $x_1^*$ & $x_2^*$ & $x_3^*$ & $x_4^*$ & $x_5^*$ & $t=KM/N$ \\ 
\hline 
0.1 & 0.5 & 0.1   & 0       & 0 & 0   & 0 & 0.5  \\
0.2 &   0 & 0.2   & 0       & 0 & 0   & 0 & 1  \\
0.3 &   0 & 0.1   & 0.05    & 0 & 0   & 0 & 1.5  \\
0.5 &   0 & 0     & 0.05    &  0.05   & 0 & 0 & 2.5  \\
0.8 &   0 & 0     & 0       & 0 & 0.2 & 0 & 4  \\
0.9 &   0 & 0     & 0       & 0 & 0.1 & 0.5 & 4.5  \\
\end{tabular}
\caption{\small{Optimal file placement parameters for $K=5$.}}\label{tab_sol}
\end{table} 

\vspace*{-.2cm}
\section{Adaptive Caching Scheme}\label{sec_ada}
We now design an adaptive delivery method that benefits from the redundancies in the user requests without changing the cache content. Further, we derive a lower bound on the delivery rate of the redundant demand vectors. 
\vspace*{-.1in}
\subsection{Adaptive Delivery Method}
For the adaptive method, we introduce  an extra step to the delivery phase, which takes place after receiving each request vector and before the transmission of the server messages to the users. In this step, the server decides whether to send each part of the requested files through the corresponding coded message in Algorithm~\ref{alg_del0d} or through an uncoded message. The use of uncoded messages instead of coded messages to deliver file $n$ is equivalent to transferring bits from $V^n_{\mathcal{S}}:s>0$ to $V^n_{\emptyset}$. Notice that by such a transfer, the cache only ignores parts of its content and it does not change the actual placement of files.  

Let $\hat{V}^n_{\mathcal{S}}$ represent the subset of the bits of file $n$ exclusively cached at $\mathcal{S}$ after the transfer is done, and 
\begin{align}
y^n_{\mathcal{S}}\triangleq |\hat{V}^n_{\mathcal{S}}|/F.
\end{align}
In our delivery method, the server first optimizes $y_\mathcal{S}^n$. Then, it arbitrarily picks $y^{n}_{\mathcal{S}}F$ bits of ${V}^{n}_{\mathcal{S}}$ to form $\hat{V}^{n}_{\mathcal{S}}$, and adds the rest of the bits to $\hat{V}^{n}_{\emptyset}$. Finally, it uses Algorithm~\ref{alg_del0d} for delivery based on the resulting subsets $\hat{V}^n_{\mathcal{S}}$ instead of ${V}^n_{\mathcal{S}}$. 

We now find the optimal lengths of the updated partition sets $\hat{V}^n_{\mathcal{S}} $ to minimize the sum of the  lengths of messages $\oplus_{k\in\mathcal{S}}\hat{V}^{d_k}_{\mathcal{S}\backslash\{k\}}$ over all the subsets $\mathcal{S}\subset \mathcal{K}$. 
Assume that the caches have requested $L\leq K$ distinct files in the current demand vector. Denote by $\mathcal{D}$, the set of the distinct files requested in the current demand vector. Note that $|\mathcal{D}|=L$, and both $\mathcal{D}$ and $L$ evolve with time.  For a fixed demand vector (the current demand vector), the rate minimization problem is given by
\begin{align}\label{eq_opt_o}
\begin{aligned}
& \underset{\substack{y^{d_k}_{\mathcal{S}}}}{\text{minimize}}
& & \sum_{\mathcal{S}:\mathcal{S}\subset\mathcal{K}}\max_{ k\in\mathcal{S}}y^{d_k}_{\mathcal{S}\backslash\{k\}}\\\
& \text{subject to}
& & \sum_{\mathcal{S}:\mathcal{S}\subset\mathcal{K}}y^{d_k}_{\mathcal{S}}=1,\quad\forall\, d_k\in\mathcal{D}\\
&&& 0\leq y^{d_k}_{\mathcal{S}} \leq x_{|\mathcal{S}|},\quad\forall\, d_k\in\mathcal{D}, \; \forall\mathcal{S}\subset\mathcal{K}:|\mathcal{S}|>0\\
&&& 0\leq y^{d_k}_{\emptyset} \leq 1,\,\qquad\forall\, d_k\in\mathcal{D}.
\end{aligned}
\end{align}
In (\ref{eq_opt_o}), $x_{|\mathcal{S}|}=|V^n_{\mathcal{S}}|$ are known from the placement phase, and are given by (\ref{eq_dec_plc}) and (\ref{eq_optplc}) for the decentralized and centralized placements, respectively.  $\max_{ k\in\mathcal{S}}y^{d_k}_{\mathcal{S}\backslash\{k\}}$ is the length of the message $\oplus_{k\in\mathcal{S}}\hat{V}^{d_k}_{\mathcal{S}\backslash\{k\}}$. Thus, the objective function is the rate of Algorithm~\ref{alg_del0d}  operating based on the adjusted  subsets $\hat{V}^n_{\mathcal{S}}$.
Similar to (\ref{eq_opt}), the equality constraint of (\ref{eq_opt_o})  is the partition constraint. Also, the constraints on the ranges of the parameters let the server to use uncoded messages instead of coded messages, but not vice versa.

Problem (\ref{eq_opt_o}) can be posed as a linear programming problem by the standard technique of defining ancillary variables  
\begin{align}
z_\mathcal{S}=\max_{ k\in\mathcal{S}} y^{d_k}_{\mathcal{S}\backslash\{k\}}
\end{align}
and adding the extra constraints 
\begin{align}
z_\mathcal{S}\geq y^{d_k}_{\mathcal{S}\backslash\{k\}} ,\;
z_\mathcal{S}\leq -y^{d_k}_{\mathcal{S}\backslash\{k\}},\quad k\in\mathcal{S}
\end{align}
for all $\mathcal{S}\in\mathcal{K}:|\mathcal{S}|>0$ \cite[Sec.~4.3]{Boyd:2004}. 
The resulting linear programming problem can be solved numerically for $y^{*d_k}_{\mathcal{S}}$. Algorithm~\ref{alg_ada} shows the adaptive delivery scheme.
\begin{algorithm}                      
\caption{\small{Original Adaptive Delivery Algorithm}}          
\label{alg_ada}                           
\begin{algorithmic}[1]                   
\REQUIRE $\{V^{n}_{\mathcal{S}}\}_{n=1,...,N,\,,\mathcal{S}\subset \mathcal{K}}$\quad\textit{\#  From the placement phase}
\STATE  \textbf{Procedure} AdaptiveDelivery($d_1,...,d_K$)\\ \vspace*{.1cm}   
	\textit{\# Message Selection Step}
		\STATE $\mathcal{D}\leftarrow \text{unique}(d_1,...,d_K)$ \# \textit{set of  distinct files requested}
	\STATE $\{y^{*\, d_k}_{\mathcal{S}}\}_{d_k\in\mathcal{D},\mathcal{S}\subset \mathcal{K}}\leftarrow$ Solution of Problem (\ref{eq_opt_o})
	\FOR{$d_k\in\mathcal{D}$}
		\STATE $\hat{V}^{d_k}_\emptyset\leftarrow  \emptyset$ \quad \textit{\#  initialization of $\hat{V}^{d_k}_\emptyset$}
    \FOR{$\mathcal{S}\subset \mathcal{K}$}
	\STATE $\hat{V}^{d_k}_\mathcal{S}\leftarrow  \{\text{first } y^{*\,d_k}_{\mathcal{S}}F \text{ bits of }{V}^{d_k}_\mathcal{S}\}$ 
		\STATE $\hat{V}^{d_k}_\emptyset\leftarrow  \hat{V}^{d_k}_\emptyset \cup \{\text{ last } (1-y^{*\,d_k}_{\mathcal{S}})F \text{ bits of } {V}^{d_k}_\mathcal{S}\}$ 
	\ENDFOR
	\ENDFOR\vspace*{.2cm}   
	
	 \textit{\# Message Construction Step}
    \FOR{$s=K,K-1,...,1$}
    \FOR{$\mathcal{S}\subset \mathcal{K}:|\mathcal{S}|=s$}
        \STATE server sends $\oplus_{k\in\mathcal{S}}\hat{V}^{d_k}_{\mathcal{S}\backslash\{k\}}$
    \ENDFOR
    \ENDFOR
\end{algorithmic}
\end{algorithm}
\subsection{Simplified Adaptive Delivery}  
A simplified version of the message selection step can be formulated by only taking the number of distinct requests $L$ into account, and ignoring the redundancy pattern of the demand vector. Then, because of the symmetry, we set $y_{\mathcal{S}}^n=y_s$ for all $n$ and all $\mathcal{S}:|\mathcal{S}|=s$. This leads to
\begin{align}\label{eq_opt2}
\begin{aligned}
& \underset{y_s}{\text{minimize}}
& & Ly_0+\sum_{s=1}^{K-1}\binom{K}{s+1}y_s\\
& \text{subject to}
& & \sum_{s=0}^{K}\binom{K}{s}y_s=1\\
&&& 0\leq y_s \leq x_s,\; s=1,...,K\\
&&& 0\leq y_0 \leq 1
\end{aligned}
\end{align}
as the simplified message selection problem. 
\begin{prop}\label{lemma_ada}
Let $\hat{s}=\lfloor\frac{K-L}{L+1}\rfloor$. Optimal parameters for the simplified message selection problem of (\ref{eq_opt2})  are given by
\begin{align}\label{eq_ada_sim}
    y_s^* = \begin{cases}
    			   \sum_{i=1,...,\hat{s}}  \binom{K}{i}x_i,              & s=0\\
               0,               & s = 1,...,\hat{s}\\
               x_s,             & s =\hat{s}+1,...,K
            \end{cases}.
\end{align}
\end{prop}
\begin{IEEEproof}
If we transfer bits from the subsets ${V}^n_{\mathcal{S}}:|\mathcal{S}|=s$ to ${V}^n_{\emptyset}$, the resulting change in the rate will be
$L\binom{K}{s}x_s-\binom{K}{s+1}x_s$.
We  transfer the bits only if this difference is negative. This is the case when $s\leq\hat{s}$. This results to the parameters  of (\ref{eq_ada_sim}).
\end{IEEEproof}
\noindent Algorithm~\ref{alg_ada_sim} shows the simplified adaptive delivery scheme.
\begin{algorithm}                      
\caption{\small{Simplified Adaptive Delivery Algorithm}}          
\label{alg_ada_sim}                           
\begin{algorithmic}[1]                   
\REQUIRE $\{V^{n}_{\mathcal{S}}\}_{ n=1,...,N,\,\mathcal{S}\subset \mathcal{K}}$ \quad \textit{\#  From the placement phase}
\STATE  \textbf{Procedure} SimplifiedAdaptiveDelivery($d_1,...,d_K$)\\ \vspace*{.1cm}   
	\textit{\# Message Selection Step}
	\STATE $L=size(unique(d_1,...,d_K))$  \textit{\# number of distinct requests}
	\STATE $\hat{s}\leftarrow\lfloor\frac{K-L}{L+1}\rfloor$ 
	\FOR{$d_k\in\mathcal{D}$}
		\STATE $\hat{V}^{d_k}_\emptyset\leftarrow  \cup_{\mathcal{S}:s\leq\hat{s}} {V}^{d_k}_\mathcal{S}$ \!\!\textit{\#\! corresponds to the first rule of (\ref{eq_ada_sim})}
    \FOR{$\mathcal{S}\subset \mathcal{K}:|\mathcal{S}|>0$}
    \IF{$|\mathcal{S}|\leq \hat{s}$}
    	\STATE $\hat{V}^{d_k}_\emptyset\leftarrow \emptyset$ \;\textit{\# corresponds to the second rule of (\ref{eq_ada_sim})}
    	\ELSE
    	\STATE $\hat{V}^{d_k}_\mathcal{S}\leftarrow {V}^{d_k}_\mathcal{S}$ \textit{\# corresponds to the third rule of (\ref{eq_ada_sim})}
    	\ENDIF
    	\ENDFOR
	\ENDFOR\vspace*{.2cm}   
	
	 \textit{\# Message Construction Step}
    \FOR{$s=K,K-1,...,1$}
    \FOR{$\mathcal{S}\subset \mathcal{K}:|\mathcal{S}|=s$}
        \STATE server sends $\oplus_{k\in\mathcal{S}}\hat{V}^{d_k}_{\mathcal{S}\backslash\{k\}}$
    \ENDFOR
    \ENDFOR
\end{algorithmic}
\end{algorithm}

\subsection{Lower Bound}
Let $R_L^*(M)$ denote the smallest rate that is achievable for every possible demand vector with $L$ distinct requests. Proposition~\ref{lem_cut} gives a lower bound on $R_L^*(M)$ based on a cutset bound argument. 
\begin{prop}[Cutset Bound]\label{lem_cut}
Assume that $K$ caches request $L\leq K$ distinct files. Then, $R_L^*(M)$ must satisfy
\begin{align}\label{eq_cut}
R_L^*(M)\geq \max_{s\in\{1,...,L\}} \left(s-\frac{s}{{\lfloor N/s \rfloor}}M\right).
\end{align}
\end{prop}
\textit{Proof.} See Appendix~\ref{app_cut}. 

\section{Numerical Examples and Simulation Results}\label{sec_sim}
In this section, we investigate the performance of the proposed adaptive delivery method through numerical examples and computer simulations. 

\subsection{Numerical Examples for Specific Demand Vectors}
\begin{figure}
\centering
\includegraphics[width=4in]{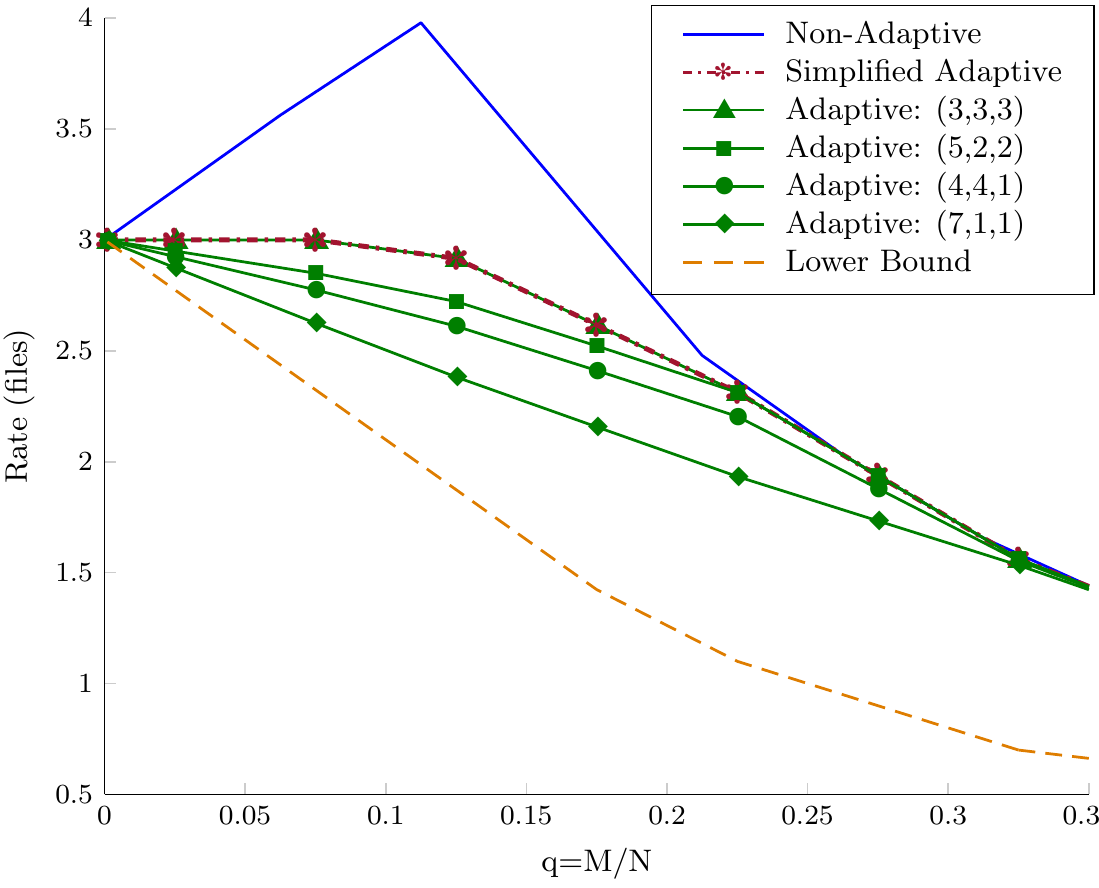}\caption{\small{Comparison of the rate of different delivery schemes for a system with $K=9$ caches. In all cases, centralized placement is used and the users only request $L=3$ distinct files. $(k_1,k_2,k_3)$ shows the number of users requesting each file.}}\label{fig_1L_cen}
\end{figure}
We first consider the performance of the adaptive methods for specific instances of the demand vector. Fig.~\ref{fig_1L_cen} shows the delivery rates of the non-adaptive delivery scheme of Algorithm~\ref{alg_del0d}, the simplified  and the original adaptive schemes, and the lower bound in Proposition~\ref{lem_cut}, for a network of $K=9$ caches.  
The  placement  in all cases is identical and is accomplished through the centralized scheme with the parameters in (\ref{eq_optplc}).  Also, we calculate the rate of the non-adaptive scheme  by (\ref{eq_rate_L}). In this example, we have considered four redundancy patterns for the demand vector, all with $L=3$ distinct file requests.  As shown in Fig.~\ref{fig_1L_cen}, the rate of the non-adaptive scheme,  the simplified adaptive scheme and the lower bound only depend on $L$ and not the specific redundancy pattern.  In contrast, the rate of the original adaptive method depends on the redundancy pattern which has led to different rates for the different patterns.  
\begin{table}
\centering
\begin{tabular}{c|c|ccccc|C{2cm}} 
Delivery  & Redundancy & \multicolumn{4}{c}{$M/N$} \\ 
\cline{3-6}
Method  &  Pattern & $0.025$ & $0.1$ & $0.15$ & $0.2$ \\ 
\hline 
Simplified Adaptive  & All      &  49\% & 52\%   & 37\%    & 13\%   \\
Adaptive             &$(3,3,3)$ &  49\% & 52\%   & 37\%    & 13\% \\
Adaptive             & $(5,2,2)$&  61\% & 61\%   & 45\%    & 17\% \\
Adaptive 		    &$(4,4,1)$  &  66\% & 66\%   & 51\%    &  25\% \\
Adaptive 			& $(7,1,1)$ &  78\% & 76\%   & 64\%    & 43\% \\
\end{tabular}
\caption{\small{Improvement of the performance gap to the lower bound in Fig.~\ref{fig_1L_cen}.}}\label{tab_per}
\end{table}
\begin{figure}
\centering
\begin{subfigure}[b]{\columnwidth}
\centering
\includegraphics[width=4in]{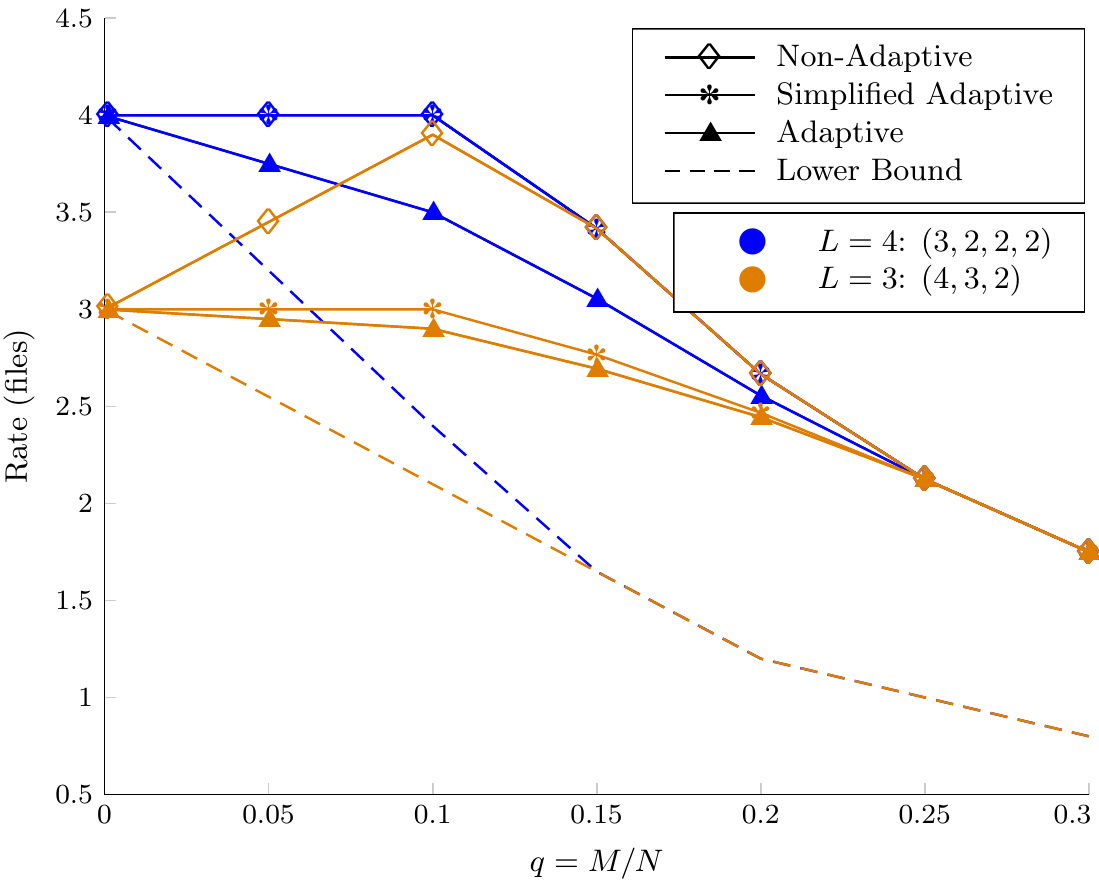}
\caption{Centralized Placement}\label{fig_2L_a}
\end{subfigure}

\begin{subfigure}[b]{\columnwidth}
\centering
\includegraphics[width=4in]{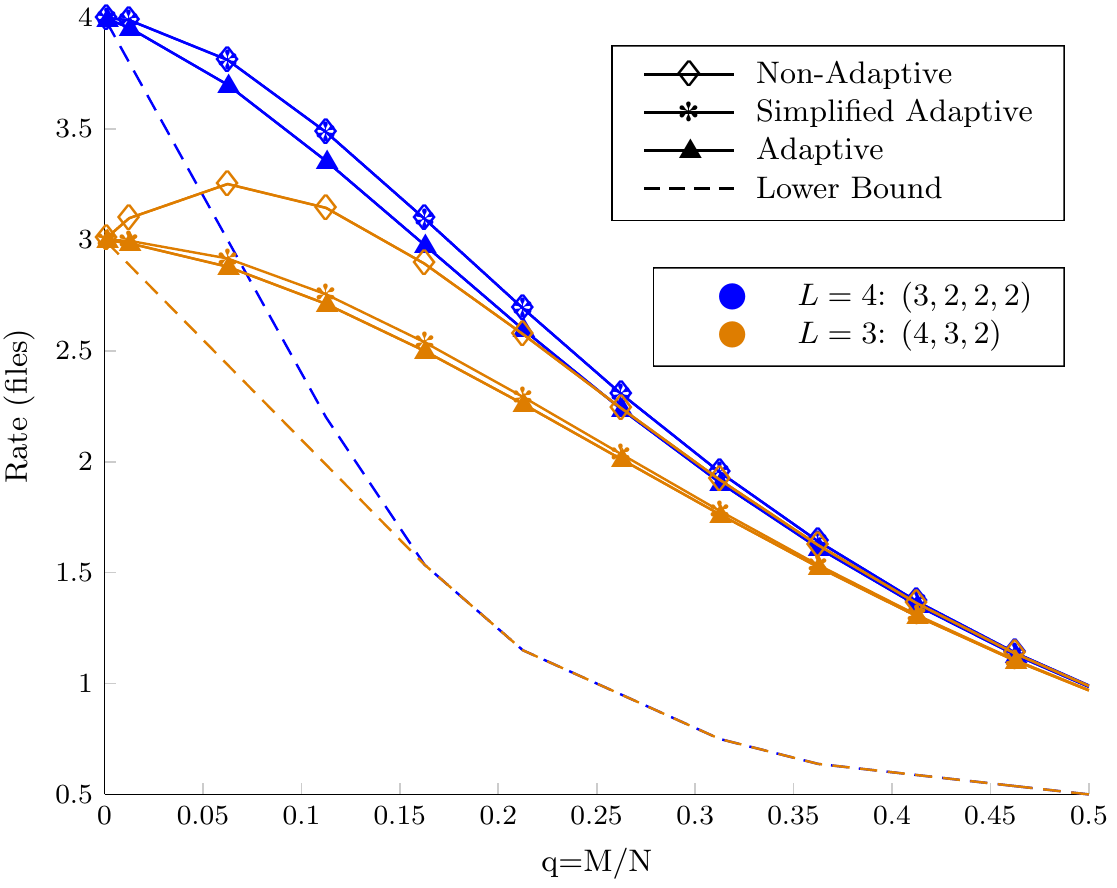}
\caption{Decentralized Placement}\label{fig_2L_b}
\end{subfigure}
\caption{\small{Comparison of the rate of different delivery schemes for a system with $K=9$ caches.}}\label{fig_2L}
\end{figure}

In Fig.~\ref{fig_1L_cen}, we observe a considerable improvement in the delivery rate for $\frac{M}{N}\leq 0.25$ when the adaptive methods are used.  Table~\ref{tab_per} shows the reduction in the gap between the non-adaptive delivery rate and the lower bound when the adaptive schemes are used. At some storage capacities, we observe 50\% and 75\% reduction in the gap for redundancy patterns  $(3,3,3)$ and $(7,1,1)$, respectively. Also,
we notice that for the symmetric redundancy pattern $(3,3,3)$, both adaptive methods led to the same delivery rate. As the redundancy pattern gets more asymmetric, the gap between the rate of the original and the simplified adaptive methods increases. 
Further, we observe that unlike the adaptive schemes, the delivery rate of the non-adaptive method increases with the storage capacity for small $M/N$. This shows the inefficiency of Algorithm~\ref{alg_del0d} to deliver the redundant requests.  
\begin{figure}
\centering
\begin{subfigure}[b]{\columnwidth}
\centering
\includegraphics[width=4in]{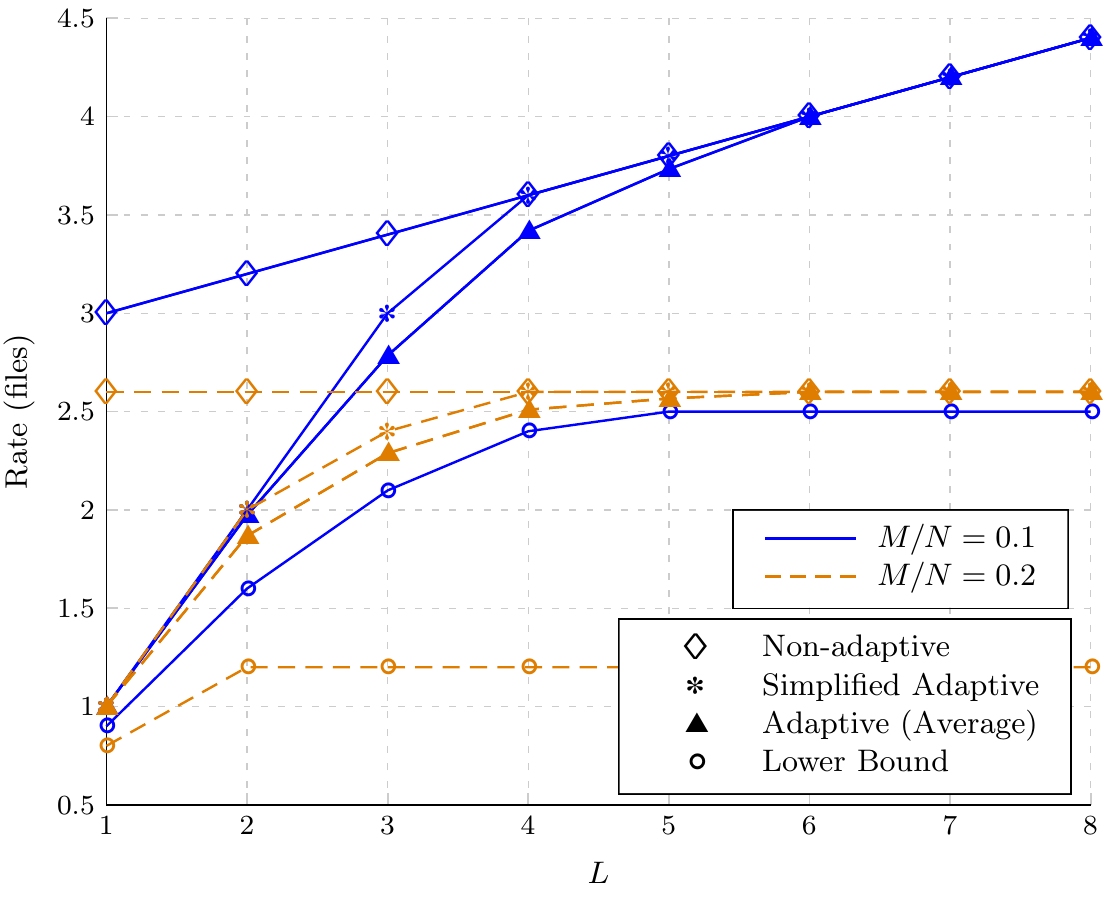}
\caption{Centralized Placement}\label{fig_allL_a}
\end{subfigure}

\begin{subfigure}[b]{\columnwidth}
\centering
\includegraphics[width=4in]{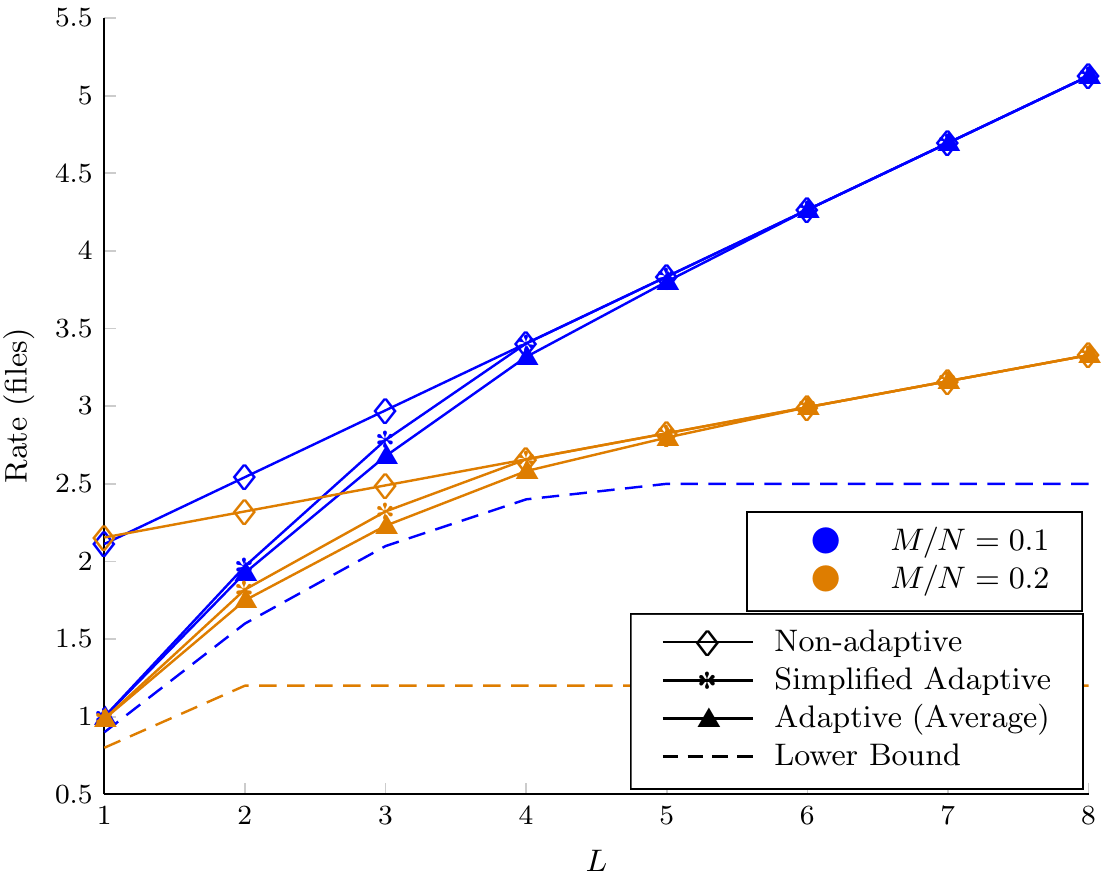}
\caption{Decentralized Placement}
\end{subfigure}\label{fig_allL_b}
\caption{\small{The effect of the number of distinct files requested on the delivery rate. Here $K=8$ and $N=10^3$.}}\label{fig_allL}
\end{figure}

Fig.~\ref{fig_2L} compares the performance of the delivery methods for two different redundancy levels $L=3$ and $L=4$. The results are shown in Figs.~\ref{fig_2L_a} and \ref{fig_2L_b} for the cases where the centralized and decentralized placement schemes are used, respectively.   For $L=4$, the rate is improved only by the original adaptive method. In general, the simplified method requires higher redundancy levels (smaller $L$) compared to the original adaptive method, to be able to improve the rate. This fact is shown in Fig.~\ref{fig_allL}, where the delivery rates are plotted versus $L$. For the original adaptive method, the delivery rate is averaged over all the redundancy patterns with $L$ distinct requests. One notices that the reduction in the delivery rate of the non-adaptive method is more considerable for smaller $M/N$. This is because when $M/N$ is small, a large number of bits are in $V_{\emptyset}^n$ subsets and need  to be delivered through uncoded messages. Based on Algorithm~\ref{alg_del0d}, the number of uncoded messages decreases by decreasing $L$. So, the reduction in the rate is larger when $M/N$ is small.

\subsection{Simulation of the Network Dynamics} 
We now investigate the average rates of the different delivery methods through a stochastic modeling of the dynamics of a caching network. 
Consider a graph representation of the network where vertices represent the caches. An (undirected) edge between two vertices shows that the requests of the corresponding caches are correlated.

To model the correlation between the requests, we assume that each cache $k$ requests a file, either based on its neighbours previous requests with probability $r$ or independently with probability $1-r$. In the former case, $k$ chooses a file from the set of the last files requested by its neighbours uniformly at random. However, when choosing independently, cache $k$ picks a file $n$ from the library of $N$ files based on the popularity distribution of the files $p_n$.  
In our simulations, we mainly use uniform popularity distribution, as it is the focus of this paper. We also consider a scenario where the file popularities are assumed to be uniform during the placement phase, but the actual demands in the delivery phase follow a non-uniform distribution. 
We use Zipf distribution with parameter $\theta$ to model the non-uniform file popularities. This gives $p_n=\frac{(1/n)^\theta}{\sum_{m=1}^N(1/m)^\theta}$ \cite{Zipf:1999}.  The larger $\theta$ is, the more non-uniform is the popularity distribution. Typical values of $\theta$ are between $0.5$ and $2$\cite{Maddah_nonuniform:2014}. $\theta=0$ corresponds to uniform distribution.

The model described above completely determines the conditional probabilities of the users' requests. The chance of requesting file $n$ by cache $k$ can be written as
\begin{align}\label{eq_conds}
\hat{p}_{n,k}=
 \begin{cases}
               \frac{1}{|\mathcal{N}(k)|}r+p_n(1-r),  & n\in\mathcal{N}(k)\\
               p_n(1-r),           				      & \text{otherwise}\\
\end{cases}
\end{align}
where $\mathcal{N}(k)$ is the set of the last files requested by the neighbour caches.  
We use Gibbs sampling \cite[Sec.~24.2]{Murphy:2012}, \cite[Sec.~3]{Fischer:2012} to generate sample vectors from the joint distribution of the user demands based on the network graph and (\ref{eq_conds}). 
In our simulations, we set $K=8$ and $N=10^3$. We assume a complete graph for the network, i.e., each vertex is of degree $K-1$. We use $r$ to control the dependency level of the users' requests. We also control the popularity distribution by $\theta$. To use Gibbs sampling, we need to 
give the underlying Markov chain enough burn-in time to reach its stationary distribution. We use the estimated potential scale reduction (ESPR) convergence criterion in \cite[Sec.~24.4.3.1]{Murphy:2012} with $5$ chains, to determine the burn-in time required. Ignoring the first $150$ sample vectors, i.e., $8\times150$ samples, suffices to get $|\text{ESPR}-1|\leq 0.01$, which shows that the stationary distribution is reached.  
We use $10^3$ sample vectors after the burn-in time to evaluate the average rate of the different delivery schemes.  
\begin{figure}
\centering
\includegraphics[width=4in]{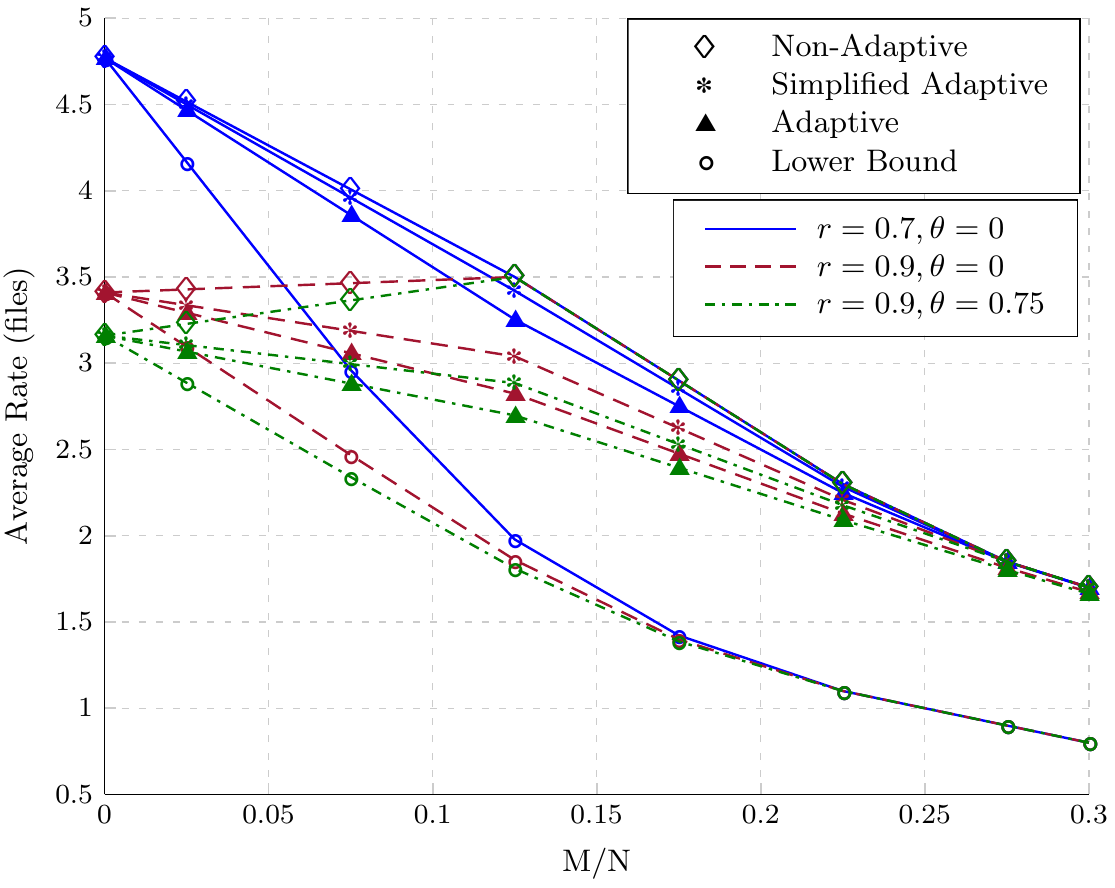}\caption{\small{Performance of the different delivery schemes in terms of their avergae delivery rates. Here, $K=8$, $N=1000$ and the central placement is used.}}\label{fig_r}
\end{figure}
\noindent Table~\ref{tab_cor} presents more details about the  correlation coefficients and the redundancy levels obtained empirically for each simulation. 
\begin{table}
\centering
\begin{tabular}{c|lllll} 
&$r=0.7$  & $r=0.9$ & $r=0.9$ \\ 
&$\theta=0$  & $\theta=0$ & $\theta=0.75$ \\ 
\hline
Maximum $\rho_{ij}$    &0.19 & 0.34& 0.34  \\
Average $\rho_{ij}$    &0.16 &0.32& 0.31  \\
Average $L$                                &4.80& 3.41& 3.18  \\
\end{tabular}
\caption{\small{Empirical correlation coefficients and the resulting number of distinct files per demand vector for the simulations in Fig.~\ref{fig_r}. $\rho_{ij}$ is the correlation coefficient between the requests of caches $i$ and $j$, $i\neq j$.}}\label{tab_cor}
\end{table}

\noindent Fig.~\ref{fig_r} shows the resulting average delivery rates.  It also shows a lower bound on the average rate that is calculated by taking the average of the lower bounds of (\ref{eq_cut}) for the sample demand vectors used. We observe that as requests become more correlated (larger $r$) and the file popularities get more non-uniform (larger $\theta$), the adaptive method makes larger improvements in the rate.
Also, the adaptive schemes are effective in decreasing the average delivery rate for $M/N<0.25$. The improvement in the performance gap to the lower bound is shown in Table~\ref{tab_per2}.

\begin{table}
\centering
a) $M/N=0.075$

\smallskip
\begin{tabular}{c|lllll} 
&$r=0.7$&  $r=0.9$ & $r=0.9$ \\
Delivery Method&$\theta=0$&  $\theta=0$ & $\theta=0.75$ \\
\hline
Adaptive             &14\%& 41\% & 47\%    \\
Simplified Adaptive  &5\% & 28\% & 36\%  \\
\end{tabular}

\medskip
b) $M/N=0.125$

\smallskip
\centering
\begin{tabular}{c|lllll} 
&$r=0.7$&  $r=0.9$ & $r=0.9$ \\
Delivery Method&$\theta=0$&  $\theta=0$ & $\theta=0.75$ \\
\hline 
Adaptive             & 16\%& 41\% & 48\%   \\
Simplified Adaptive& 5\% &  28\% & 36\%   \\
\end{tabular}
\caption{\small{Improvement in the performance gap to the lower bound in Fig.~\ref{fig_r}.}}\label{tab_per2}
\end{table}

\section{Conclusion}\label{sec_con}
We proposed a new delivery scheme for caching networks that exploits the redundancies in the users' demand vector to reduce the delivery traffic. The proposed scheme allows the server to decide between the use of coded messages of \cite{Maddah_decentralized:2014} or uncoded messages for the delivery of each part of the files requested. This choice is made based on the redundancy pattern of the requests in the current demand vector. The server's decision making process is formulated as a linear programming problem which must be solved numerically. To facilitate the decision making process, a simplified decision rule is also derived analytically.  Further, we derived a lower bound on the delivery rate of redundant demands based on a cutset bound argument. The proposed adaptive schemes are shown to significantly improve the delivery rate through several numerical examples. They decreased the performance gap of the non-adaptive method to the lower bound by up to 80\% for highly redundant demand vectors. We also investigated the dynamics of a caching network through Markov chain Monte-Carlo simulations and reported the average delivery rate of the adaptive schemes. The adaptive methods considerably outperform the non-adaptive methods in terms of the resulting average delivery rates when $M/N<0.25$. We also generalized the application of the centralized placement scheme of \cite{Maddah_limits:2014} to caching networks with non-integer $KM/N$.

\appendices
\section{Proof of Proposition~\ref{lem_plc}}\label{app_plc}
\begin{IEEEproof} By the Karush-Kuhn-Tucker (KKT) conditions \cite[Sec.~5.5.3]{Boyd:2004} for optimization problem (\ref{eq_opt}), we get
\begin{small}
\begin{align}\label{eq_KKT1}
\binom{K}{s+1}-\lambda_s+\binom{K-1}{s-1}\lambda'+\nu\binom{K}{s}=0,\;s=0,...,K
\end{align}
\end{small}
where $\lambda_s\geq 0$ is the Lagrange multiplier for the inequality constraint $x_s\geq 0$, and $\lambda'\geq 0$ and $\nu$ are the Lagrange multipliers for the capacity inequality constraint and the partition equality constraint, respectively.
 
KKT conditions require $\lambda_s x_s^*=0$. As a result,  $x^*_s>0$ requires $\lambda_s=0$. From (\ref{eq_KKT1}), we can have $\lambda_s=0$ for at most two $s$ indices. This is because  $\lambda'$ and $\nu$ provide only two degrees of freedom to set $\binom{K}{s+1}+\binom{K-1}{s-1}\lambda'+\nu\binom{K}{s}=0$ and the coefficient matrix given by these linear equations is rank 2. As a result, we have either one or two $x^*_s$ values greater than zero. We consider each case separately.

First, assume that only $x^*_{s_0}$ is non-zero. Then, the equality constraint and the capacity constraint require $x^*_{s_0}=1/\binom{K}{s_0}$ and $s_0=K\frac{M}{N}$, respectively. If $t=K\frac{M}{N}$ is an integer, the optimal solution is achieved, as is given by (\ref{eq_optplc_1x}).  Otherwise, not all the storage capacity is used and the solution is not optimal, i.e., the optimal solution has more than one non-zero $x^*_s$. So, as the second case, let $x^*_s>0$ for exactly two values of $s$, namely $s_1$ and $s_2>s_1$. From the storage and partition constraints we get 
\begin{subequations}\label{eq_xs}
\begin{align}
x^*_{s_1}=\frac{\binom{K-1}{s_2-1}-\binom{K}{s2}M/N}
{\binom{K}{s2}\binom{K-1}{s_1-1}-\binom{K}{s1}\binom{K-1}{s2-1}}\\
x^*_{s_2}=\frac{\binom{K-1}{s_1-1}-\binom{K}{s1}M/N}
{\binom{K}{s2}\binom{K-1}{s_1-1}-\binom{K}{s1}\binom{K-1}{s2-1}}.
\end{align}
\end{subequations}
Since $x^*_{s_1},x^*_{s_2}>0$, (\ref{eq_xs}) requires
\begin{align}\label{eq_scon}
s_1\leq K\frac{M}{N},\quad s_2\geq K\frac{M}{N}.
\end{align}
Given (\ref{eq_xs}), the objective function  
$\binom{K}{s_1+1}x^*_{s_1}+\binom{K}{s_2+1}x^*_{s_2}$
simplifies to 
\begin{align}\label{eq_last}
K-(K+1)\frac{KM/N+s_1s_2}{(s_1+1)(s_2+1)}.
\end{align}
The function in (\ref{eq_last}) is decreasing in $s_1$ and increasing in $s_2$ in the region specified by (\ref{eq_scon}). Therefore, to minimize the objective function, $s_1$ must take its largest value $\lfloor K\frac{M}{N}\rfloor$, and $s_2$ must take its smallest value $\lceil K\frac{M}{N} \rceil$. Substitution of these values in (\ref{eq_xs}) gives the optimal parameters in (\ref{eq_optplc_2x}). This completes the proof of Proposition~\ref{lem_plc}.
\end{IEEEproof}

\section{Proof of Proposition~\ref{lem_cut}}\label{app_cut}
\begin{IEEEproof} We modify the cutset bound argument of \cite[Sec.~VI]{Maddah_limits:2014} to bound the minimum delivery rate of the demand vectors with $L\leq K$ distinct requests. 

Let $\mathcal{S}$ be a subset of caches with $|\mathcal{S}|=s$, such that there are no two caches in $\mathcal{S}$ with identical user requests. Assume that these caches have requested files $1,...\, ,s$ from the library of $N$ files. Let $X_1$ denote the server's input to the shared link which determines files $1,..,s$. Similarly, assume that the same users request files $(i-1)s+1,...,is$ and the server input $X_i$ determines the files requested. Let $i=1,...,\lfloor N/s\rfloor$. 

Consider the cut separating $X_1,...,X_{\lfloor N/s \rfloor}$ and the caches in $\mathcal{S}$ from the corresponding users (see Fig.~\ref{fig_cutset}). Since we assume that the coded caching scheme works and all files are perfectly decoded, the total information available to the users in the cut should be more than or equal to the total information requested by them. In other words, 
\begin{align}\nonumber
{\lfloor N/s \rfloor}R_L^*(M)+sM\geq s{\lfloor N/s \rfloor}.
\end{align}
Since $s$ can accept any value between $1$ and $L$, (\ref{eq_cut}) results.
\end{IEEEproof}
\definecolor{dimgray}{rgb}{0.41, 0.41, 0.41}
\definecolor{darkgray}{rgb}{0.66, 0.66, 0.66}
\definecolor{lightgray}{rgb}{0.83, 0.83, 0.83}
  \tikzset{
  treenode/.style = {align=center, inner sep=0pt,text centered },
  arn_r/.style = {treenode,rectangle,draw=black,
        text centered, text=black, text width=3.5em, text height=1.5em,text depth=2.15em},
 arn_ser/.style = {treenode,rectangle,draw=black,
        text centered, text=black, text width=7em, text height=1.5em,text depth=1.5em},      
  arn_c/.style = {treenode, circle, black,draw=black,
    fill=white, text width=3em},
  arn_x/.style = {treenode, rectangle, draw=black,
    minimum width=0.0em, minimum height=0.0em}
}
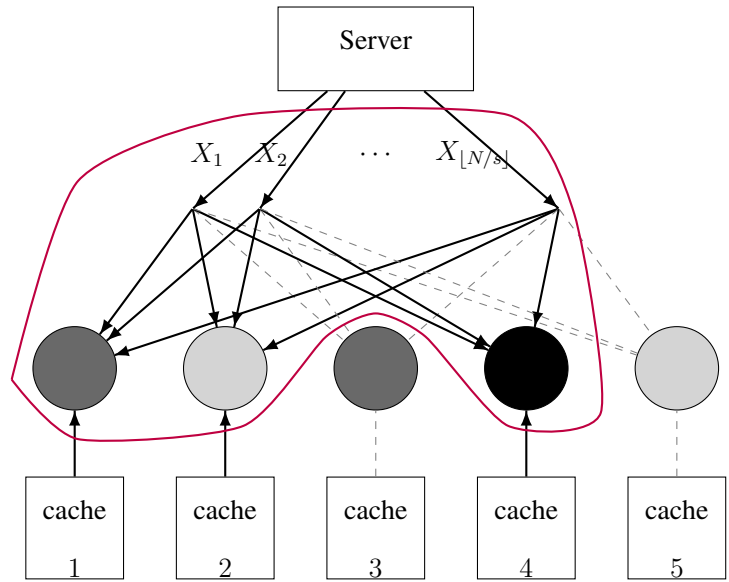
\begin{figure}
\centering
\begin{tikzpicture}[scale=.88,transform shape]
\tikzset{myptr/.style={decoration={markings,mark=at position 1 with %
    {\arrow[scale=1.1,>=latex]{>}}},postaction={decorate}},
    myptr2/.style={decoration={markings,mark=at position 1 with %
    {\arrow[scale=1.1,>=latex]{>}}},postaction={decorate}}}

\node [arn_ser] (SS) {Server}
      node [arn_r,below= of SS,draw=none] (S3) {}
            { node [below= of S3,arn_c,fill=dimgray] (c)  {} 
                        {node[below= of c,,arn_r] (cc){cache \\${3}$}}
            }
            { node [left= of c,arn_c,fill=lightgray](b) {} 
                        {node[below= of b,arn_r] (bb) {cache \\$2$}}
            }
            { node [left= of b,arn_c,fill=dimgray](a) {} 
                        {node[below= of a,arn_r] (aa) {cache \\$1$}}
            }            
            { node [right= of c,arn_c,fill=black] (d) {}
                        {node[below= of d,arn_r] (dd) {cache \\${4}$}}
            }                            
            { node [right= of d,arn_c,fill=lightgray] (e) {}
                        {node[below= of e,arn_r] (ee) {cache \\${5}$}}
            }   
            node [arn_x,left= of S3] (S2) {}
            node [arn_x,left= of S2] (S1) {}
            node [arn_x,right= of S3] (S4) {}
            node [arn_x,right= of S4] (S5) {}
;      
\node[text width=1cm,draw=none] at (-2.3,-1.6) {$X_1$};
\node[text width=1cm,draw=none] at (-1.33,-1.6) {$X_2$};
\node[text width=1cm,draw=none] at (0.25,-1.6) {$\cdots$};
\node[text width=1cm,draw=none] at (1.4,-1.6) {$X_{\lfloor N/s\rfloor}$};

\draw[draw=black,thick,myptr] (SS) --(S1);
\draw[draw=black,thick,myptr] (S1) --(a);
\draw[draw=black,thick,myptr] (S1) --(b);
\path[draw=gray,dashed] (S1) edge(c);
\draw[draw=black,thick,myptr] (S1) --(d);
\path[draw=gray,dashed] (S1) edge(e);

\draw[draw=black,thick,myptr] (SS) --(S5);
\draw[draw=black,thick,myptr] (S5) --(a);
\draw[draw=black,thick,myptr] (S5) --(b);
\path[draw=gray,dashed] (S5) edge(c);
\draw[draw=black,thick,myptr] (S5) --(d);
\path[draw=gray,dashed] (S5) edge(e);

\draw[draw=black,thick,myptr] (SS) --(S2);
\draw[draw=black,thick,myptr] (S2) --(a);
\draw[draw=black,thick,myptr] (S2) --(b);
\path[draw=gray,dashed] (S2) edge(c);
\draw[draw=black,thick,myptr] (S2) --(d);
\path[draw=gray,dashed] (S2) edge(e);

\draw[draw=black,thick,myptr2] (aa) --(a) ;
\draw[draw=black,thick,myptr2] (bb) --(b);
\path[draw=gray,dashed] (cc) edge(c);
\draw[draw=black,thick,myptr2] (dd) --(d);
\path[draw=gray,dashed] (ee) edge(e);    

            \draw[thick,purple]  plot[smooth, tension=.4] coordinates {(-5.5,-5) (-4.5,-2) (-2,-1) (2,-1) (3,-2.5) (3.4,-5.5) (2,-5.7) (0.75,-4.35) (0,-4) (-0.75,-4.35) (-2,-5.7) (-4.5,-5.9) (-5.5,-5)};                   
\end{tikzpicture}
\vspace*{.1cm}
\caption{\small{An example of a cutset separating the caches and the server from the users of the caches in $\mathcal{S}=\{1,2,4\}$. Solid (dashed) lines represent the information flow to the users selected (not selected) in the cutset. Here $s=3$ and there are ${\lfloor\frac{N}{s}\rfloor}$ server messages. Users with the same color have identical requests. Notice that no two users with the same request are picked in $\mathcal{S}$. Here $K=5$.}
}\label{fig_cutset}
\end{figure}


\bibliographystyle{ieeetr}
\bibliography{Caching.bib}

\end{document}